\documentclass{article}
\usepackage{spconf,amsmath,graphicx}
\usepackage{xcolor}
\usepackage[ruled,vlined]{algorithm2e}
\usepackage[normalem]{ulem}
\usepackage{amssymb}

\usepackage{tabularx}
\usepackage{multirow}
\usepackage{booktabs}
\usepackage{subcaption}
\usepackage{enumitem}
\usepackage{float}
\usepackage{hyperref}
\usepackage{comment}
\hypersetup{pagebackref=true,breaklinks=true,colorlinks,bookmarks=false}

\SetCommentSty{mycommfont}

\newcommand{\steph}[1]{\textcolor{blue}{Steph: #1}}

\title{A Hybrid Deep Animation Codec for low-bitrate video conferencing}
\name{Goluck Konuko$^{\dagger}$, St\'ephane Lathuili\`ere$^{\ddagger}$ Giuseppe Valenzise$^{\dagger}$}
\address{$^{\dagger}$ Université Paris-Saclay, CNRS, CentraleSupélec,  Laboratoire des signaux et systèmes\\ 
91190, Gif-sur-Yvette, France\\
$^{\ddagger}$ LTCI, Télécom Paris, Institut Polytechnique de Paris, France}
\begin{document}
%
\maketitle
\begin{abstract}
Deep generative models, and particularly facial animation schemes, can be used in video conferencing applications to efficiently compress a video through a sparse set of keypoints, without the need to transmit dense motion vectors. While these schemes bring significant coding gains over conventional video codecs at low bitrates, their performance saturates quickly when the available bandwidth increases.  In this paper, we propose a layered, hybrid coding scheme to overcome this limitation. Specifically, we extend a codec based on facial animation by adding an auxiliary stream consisting of a very low bitrate version of the video, obtained through a conventional video codec (e.g., HEVC). The animated and auxiliary videos are combined through a novel fusion module. Our results show consistent average BD-Rate gains in excess of -30\% on a large dataset of video conferencing sequences, extending the operational range of bitrates of a facial animation codec alone. 
\end{abstract}
\section{Introduction}
\label{sec:intro}

In the pursuit of higher video compression performance at low bitrates, deep generative models have been recently employed in image and video compression to overcome the intrinsic limitations of traditional video coding schemes in modeling complex pixel dependencies \cite{pad2013optimality,wang2004image}.
In particular, for video conferencing applications, we have shown in our previously proposed Deep Animation Codec (DAC)~\cite{konuko2020dac} that it is possible to code videos with talking heads at bitrates as low as 5 kbps, and reconstruct them with good visual quality, using a face animation scheme \cite{siarohin2019first}. 

Face animation employs a set of sparse keypoints to encode the motion of faces, instead of dense motion vectors used in traditional pixel-based codecs. At the decoder side, the keypoints are used to synthesize a dense motion field, which is then employed to warp a source frame (coded as an Intra frame with a conventional codec), producing a realistic, high-quality synthesized picture. A similar scheme has also been applied in concurrent or later works~\cite{wand2021conferencing,oquab2021chat}.

A main shortcoming of face animation codecs is that their performance tends to saturate quickly when the available bandwidth increases, limiting the achievable video reconstruction quality. This is partly due to the open loop coding structure, which makes it difficult to model long-term pixel dependencies as well as significant displacements in the background, introducing a quality drift. A possible solution consists in adding Intra-refresh frames to reset this drift periodically~\cite{konuko2020dac}. However, this entails a higher bitrate, which makes this option noncompetitive compared with conventional codecs. 

In this work, we solve the problem of long-term dependencies and background motion by an alternative, hybrid approach. We augment the DAC bitstream with an auxiliary stream, obtained by compressing at very low quality the original video, using a state-of-the-art conventional codec (in our work, we use HEVC). The DAC output and auxiliary decoded video are fed into a novel fusion module, which combines them to reconstruct the final result. Despite its very low bitrate (and the consequent poor visual quality), the auxiliary video provides enough information to regularize the animation and compensate for the motion estimation errors in the DAC bitstream. Using this Hybrid Deep Animation Codec (H-DAC), we obtain BDRate reductions over HEVC in excess of 30\%, and similar performance to VVC, on two test datasets composed by video conferencing sequences.

\section{Related Work}
\label{sec:related}

The application of deep learning to image and video compression has received a great deal of attention in the past few years~\cite{rippel2017real,balle2018variational,agustsson2019generative}, thanks to its ability to represent complex pixel dependencies and obtain good visual quality at low bitrate~\cite{valenzise2018quality}. Deep neural networks can be applied either to replace/improve specific coding tools (e.g., spatial prediction~\cite{wang2019enhancing}), or to optimize the whole coding pipeline in an end-to-end fashion, typically using variational auto-encoders~\cite{balle2016end,lu2019dvc}. This work falls in the former category, as we employ a deep neural network to optimize the motion prediction and compensation through face animation, but we rely on conventional entropy coding, intra-frame prediction and an auxiliary standard video stream to reconstruct the decoded video.

Deep generative models are a family of methods that aim at learning (or sampling from) the data distribution~\cite{goodfellow2015gans}. They have been used in video compression to code a picture/video at very low bitrate, e.g., by hallucinating parts of the video outside the region of interest~\cite{kaplanyan2019deepfovea}. In this work, we focus instead on synthesizing the foreground pixels of a talking head in video conferencing. To this end, we employ image animation models, which are a specific kind of generative models able to produce realistic, high-quality videos of moving faces~\cite{kim2018deep}. A typical application consists of transferring the movements of a driving sequence to a source frame~\cite{chan2018dance,Siarohin_2019_CVPR} in order to swap faces in videos and produce deep fakes~\cite{yang2019exposing}.

More recently, image animation models have been employed in video coding to enable ultra-low bitrate video conferencing~\cite{konuko2020dac,wand2021conferencing,oquab2020low, chen2022beyondkp}. In these works, the face motion is represented by a set of sparse keypoints, which are encoded and transmitted as bitstream. At the decoder side, the received keypoints are used to reconstruct a dense optical flow, which is then used to warp a source frame (intra coded) and produce an estimate of the reconstructed frame. Among these proposed methods, only the Deep Animation Codec (DAC) in~\cite{konuko2020dac} offers the possibility to vary the bitrate and quality to a certain extent, by modulating the frequency of intra refresh. In this work, we build on the basic architecture of DAC, but we extend it with an additional auxiliary stream to handle motion in the background and long-range temporal dependencies, and to increase its operational range of bitrates and qualities.

\section{Proposed coding method (H-DAC)}
\label{sec:method}
Our goal at test time is to compress an input video sequence $F_0,\hdots,F_t$ corresponding to a video conferencing scenario. To this aim, we consider that at training time, we have at our disposal many videos containing talking faces. 

\begin{figure*}[t]
	\begin{center}
		\includegraphics[width=\linewidth]{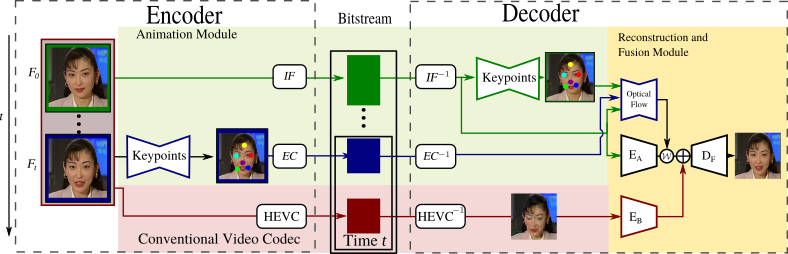}
	\end{center}
    \vspace{-0.5cm}
	\caption{Basic scheme of our proposed hybrid codec. The bitstream is composed by the low-bitrate HEVC video, the Intra frame (IF) and the entropy-coded (EC) keypoints.  The animation module employs IF, encoded by a traditional codec (e.g. HEVC Intra), to animate the subsequent inter-frames, using a set of sparse keypoints to encode motion. The fusion module concatenates the warped ($\mathcal{W}$) intra frame features and the low-quality HEVC frame features, and decodes a predicted output frame.
	}
	\label{fig:pipeline}
\end{figure*}

Our codec is divided in three main modules illustrated in Figure~\ref{fig:pipeline}. First a \textit{conventional video codec} (light red module) is used on both encoder and decoder sides to transmit the input video with a very low bitrate. In this module, any conventional video codec can be employed (in our experiments, we adopt HEVC with a low-delay configuration).

In the second module, an image animation model (green module) is employed. This module is based on the \textit{Deep Animation Codec} (DAC) proposed in our previous work~\cite{konuko2020dac}. It provides to the decoder the initial frame of the video and keypoints that are learned in an unsupervised manner in order to describe the motion between the initial frame and the current frame. Our image animation model is  described in Section~\ref{sec:imageAnimation}. Note that, at every time step $t$, the current frame is encoded via the current HEVC P-frame and the keypoint estimated in the current image $F_{t}$. 

Finally, a \emph{fusion module} combines the low-quality video provided by the conventional video codec with the output of the image animation module. The fusion and image animation modules are jointly learned by minimizing a reconstruction loss on the decoder side. Practically we employ a combination of a perceptual and adversarial losses as in~\cite{Siarohin_2019_CVPR}. These three modules are detailed in the following.

\subsection{Animation Module}
\label{sec:imageAnimation}

Image animation is based on the idea that a source image depicting a single object such as human faces can be employed to create a video sequence by transferring motion encoded as the displacement of keypoints. In our video compression context, the image animation framework is used as follows: the initial frame, interchangeably referred to as the source frame or intra frame in a group of pictures is transmitted to the decoder and all the other frames are encoded via the displacement of keypoints estimated independently in every frame.

\noindent \textbf{Intra frame compression.} The intra frame is compressed  by any  off-the-shelf codec. We apply the BPG codec here, which is equivalent to HEVC Intra. 
  The intra frame used in the animation process is the decoded frame received from the BPG codec.
 
\noindent \textbf{Keypoint extraction, quantization and coding.} As in \cite{konuko2020dac}, a keypoint detection network is used to predict the location of the keypoints in every frame. Each keypoint is associated with a $2\times 2$ Jacobian matrix as described by~\cite{Siarohin_2019_CVPR}, which describes the orientation of the local motion vectors. Dense optical flow maps are generated through a first-order Taylor expansion around the sparse keypoint locations. The estimated optical flow map is used to deform the features of the source frame before further refinement is applied. As in \cite{konuko2020dac}, we use 10 facial keypoints. The intra frame keypoints are used as reference since they are available both at the encoder and decoder side without the need for transmission. This enables an efficient 8-bit quantization process that greatly reduces the bitrate contribution of the motion keypoints. The quantized keypoint vectors are entropy-coded using a standard arithmetic codec.

\subsection{Reconstruction and Fusion Module}
The animation-based codec alone cannot handle well background motion and disocclusions. Conversely, the HEVC base layer provides poor texture details but can encode changes in the background or object disocclusions while being robust to complex non-rigid motions (e.g., hair movements) that are not captured by keypoints. In this section we describe our reconstruction and fusion module that leverages both streams to reconstruct a good quality frame at the decoder side.

The reconstruction and fusion module is depicted in the right yellow box in Figure~\ref{fig:pipeline}. It processes the outputs of the baseline video codec and the animation module. It is composed by 5 blocks. The optical flow encoder-decoder network produces a dense optical flow between the intra frame and the current one based on the decoded keypoints. The convolutional encoders $E_A$ and $E_B$ extract features from the intra frame and the base layer decoded frame, respectively. The warping operator $\mathcal{W}$~\cite{siarohin2019first} applies motion compensation in the feature domain. The warped features from the animated frame and those obtained by $E_B$ are concatenated, and the decoder $D_F$ produces the final reconstructed frame after feature fusion. 
The source code for our coding framework is available at \url{goluck-konuko.github.io}.

\section{Experiments and Results}
\label{sec:expe}

\begin{figure*}[h]
\begin{center}

\vspace{-0.3cm}
\footnotesize
\def\arraystretch{0.1}
\setlength\tabcolsep{1.1pt}
\begin{tabular}{ccccc}
Ground Truth&VVC(20Kbps)&HEVC(~20kbps)&\textbf{DAC(~20kbps)}&\textbf{H-DAC(~20kbps)}\\
\includegraphics[trim=0 0 1025 0,clip,width=0.18\textwidth]{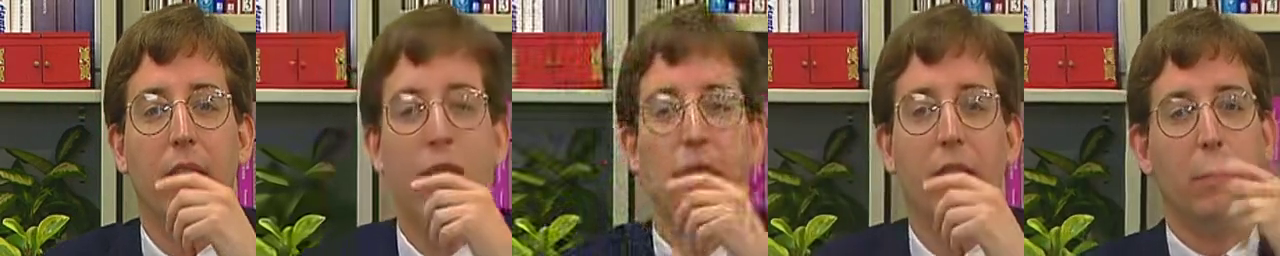}&
\includegraphics[trim=255 0 770 0,clip,width=0.18\textwidth]{figures/comparison_20_kbps/clip_224_50.png}&
\includegraphics[trim=510 0 515 0,clip,width=0.18\textwidth]{figures/comparison_20_kbps/clip_224_50.png}
&
\includegraphics[trim=1025 0 0 0,clip,width=0.18\textwidth]{figures/comparison_20_kbps/clip_224_50.png}
&\includegraphics[trim=765 0 260 0,clip,width=0.18\textwidth]{figures/comparison_20_kbps/clip_224_50.png}
\\
\includegraphics[trim=0 0 1025 0,clip,width=0.18\textwidth]{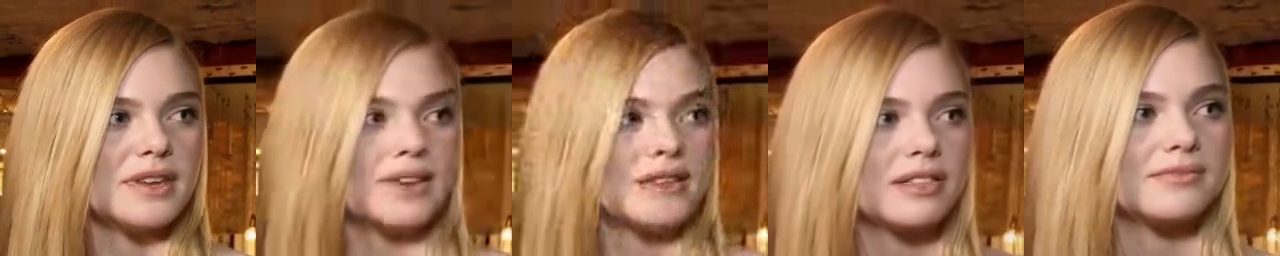}
&
\includegraphics[trim=510 0 515 0,clip,width=0.18\textwidth]{figures/comparison_20_kbps/65_50.png}
&
\includegraphics[trim=255 0 770 0,clip,width=0.18\textwidth]{figures/comparison_20_kbps/65_50.png}
&\includegraphics[trim=765 0 260 0,clip,width=0.18\textwidth]{figures/comparison_20_kbps/65_50.png}&
\includegraphics[trim=1025 0 0 0,clip,width=0.18\textwidth]{figures/comparison_20_kbps/65_50.png}
\\
\includegraphics[trim=0 0 1025 0,clip,width=0.18\textwidth]{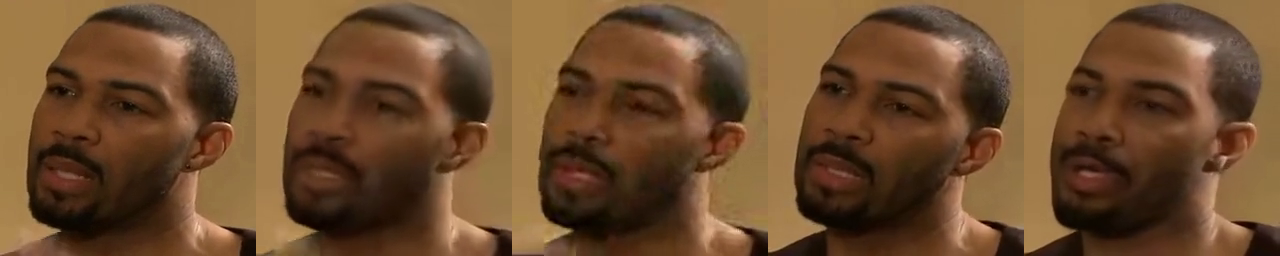}&
\includegraphics[trim=255 0 770 0,clip,width=0.18\textwidth]{figures/comparison_20_kbps/66_50.png}&
\includegraphics[trim=510 0 515 0,clip,width=0.18\textwidth]{figures/comparison_20_kbps/66_50.png}
&
\includegraphics[trim=1025 0 0 0,clip,width=0.18\textwidth]{figures/comparison_20_kbps/66_50.png}
&\includegraphics[trim=765 0 260 0,clip,width=0.18\textwidth]{figures/comparison_20_kbps/66_50.png}
\end{tabular}
\end{center}
\caption{Qualitative comparison of reconstructed images with our codec and state-of-the-art codecs at a similar bitrate. 
H-DAC significantly improves DAC in reproducing face expressions with good fidelity (e.g., the open mouth in the second row and the teeth in the third row), and is robust to non-face objects such as the hand in the first row. In addition, the synthesized images display lower distortion than HEVC and VVC.}
\vspace{-0.5cm}
\label{fig:subjective_eval}
\end{figure*}

\textbf{Datasets.}
Following \cite{konuko2020dac}, we use two datasets in our experiments. The \textbf{VoxCeleb2} dataset is a large audio-visual dataset of talking heads with about 22k videos extracted from Youtube at a resolution of 256x256 pixels~\cite{Siarohin_2019_CVPR}. 
90 sequences with complex motion patterns from the VoxCeleb2 test set are used for testing the compression framework. For more technical details, please refer to \cite{konuko2020dac}.
In addition, the \textbf{Xiph.org}~\footnote{\url{https://media.xiph.org/video/derf/}} dataset is also used as test dataset. In this case, we employ the model trained on VoxCeleb2. We downloaded the videos from Xiph.org talking humans ~\cite{konuko2020dac} and selected 16 sequences from which we crop the region of interest around the human face at a resolution of 256x256.

\noindent\textbf{Implementation details.} Regarding the network architecture, we use the motion transfer network architecture described by ~\cite{siarohin2019first} and design a fusion module with spatial attention layers from \cite{cheng2020attention}. The HEVC base layer is coded with a fixed QP value of 50 for all sequences. The network is trained in an end-to-end fashion for 100 epochs.

\noindent\textbf{Metrics.} In addition to the widely used PSNR and SSIM metrics, we adopt the msVGG loss that is the multi-scale LPIPS loss used in~\cite{siarohin2019first}.

\noindent\textbf{Comparison to state-of-the-art video codecs.} We evaluate the coding framework performance with respect to the HEVC codec under a low-delay configuration. 
Qualitative comparison are reported as Bjontengaard-Delta rate over HEVC in Table~\ref{tab:Comp-HEVC}. We observe that we obtain better performance in terms of BD rate for the three metrics and the two datasets. The gain in BD rate is especially clear in terms of msVGG since this metric is employed as training loss.

\begin{table}[t]
\begin{center}
\caption{Bjontengaard-Delta Performance of H-DAC over HEVC}
\vspace{-0.3cm}
\begin{tabular}{lccc}
\toprule
&\textbf{VoxCeleb}&\textbf{Xiph.org}\\
&{\footnotesize BD quality~/~BD rate}&\footnotesize BD quality~/~BD rate\\
\midrule
\textbf{PSNR$\uparrow$}& 1.07~/~-33.36&0.97~/~-30.7\\
\textbf{SSIM$\uparrow$}&  ~0.02/~-33.41 &  0.02~/~-28.33 \\
\textbf{msVGG $\downarrow$} & -19.16~/~-48.84 &-20.04~/~-41.64\\
  \bottomrule
\end{tabular}
\label{tab:Comp-HEVC}
\end{center}
\vspace{-0.5cm}
\end{table}

\begin{figure}[t]
\footnotesize
\begin{minipage}[c]{0.5\textwidth}
 \includegraphics[width=0.9\textwidth]{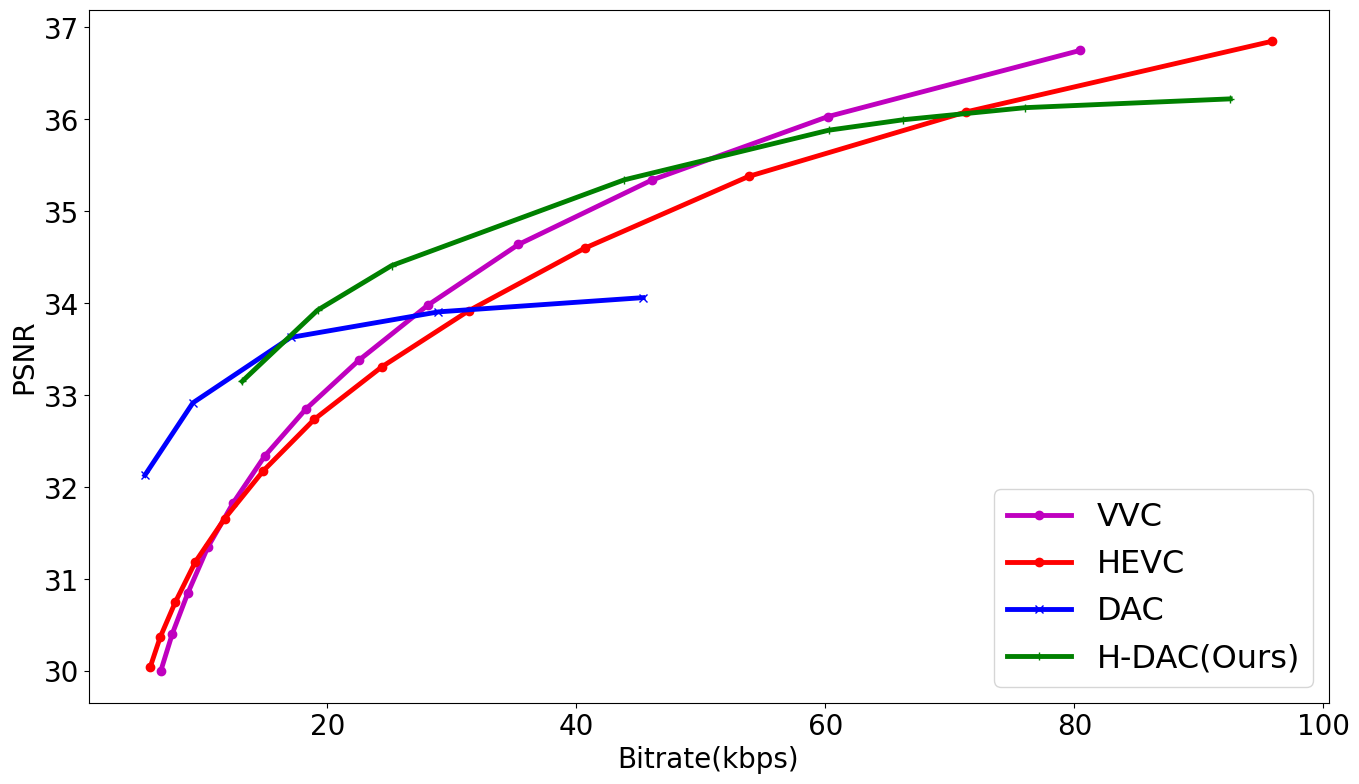}
\end{minipage}
\vspace{-0.4cm}
\caption {RD curve for 105 test videos sampled from VoxCeleb and Xiph.org test sets. The DAC is our coding framework proposed in \cite{konuko2020dac} which uses adaptive refresh for quality scalability. }
\label{fig:RD_HEVC}
\end{figure}

We show the RD curves using PSNR in Figure~\ref{fig:RD_HEVC}, where we also report the VVC performance for reference. We see that DAC~\cite{konuko2020dac} achieves very low bitrate, but its quality saturates quickly and cannot reach PSNR values above 34dB even when the bitrate increases. Conversely,
the use of a conventional video codec stream shifts the range of bitrates over which our framework can operate before reaching a saturation point. Regarding the comparison with HEVC, H-DAC consistently obtains better scores at all the considered bitrates. Interestingly, our approach achieves close performance to VVC (which has a significantly higher complexity than HEVC). Notice that we did employ only HEVC in H-DAC; however, the conventional video coding module and the intra frames in the animation module might be replaced with VVC, leading to further coding gains. 

To conclude this section, Figure~\ref{fig:subjective_eval} shows some qualitative compression results. At comparable bitrates, we observe a considerable difference in reconstruction quality, especially in scenes where we may have other elements than just faces, such as the speaker's hand within the target frame in the first row. While the conventional codecs display significant blur (VVC) and blocking (HEVC) artifacts at this bitrate, H-DAC does not produce these distortions and synthesizes images with better perceptual quality. In particular, the resulting visual quality is better than VVC despite the RD curves computed in Figure~\ref{fig:RD_HEVC}. However, a more rigorous subjective study needs to be conducted to confirm this observation, and is left for future work.

\section{Conclusion}
\label{sec:conclusion}

Using deep facial animation for compression in video conferencing applications leads to significant coding gains at ultra-low bitrates. However, the performance tends to saturate quickly as the available bandwidth increases, due to a poor capability of animation schemes in handling long-term temporal dependencies, disocclusions and background changes. In this paper, we propose H-DAC, a Hybrid Deep Animation Codec that combines face animation with an auxiliary, conventional bitstream at very low bitrate. Our results demonstrate that this approach can overcome the limitations of codecs using purely face animation to synthesize frames, bringing significant quality gains compared to state-of-the-art video codecs. 

\textbf{Acknowledgement:} This work was funded by Labex DigiCosme - Universit\'e Paris-Saclay

\bibliographystyle{IEEEbib}
\bibliography{strings,refs}

\end{document}